\documentclass[pra,superscriptaddress,showpacs,amsmath,amssymb]{revtex4}
\usepackage{graphicx}
\usepackage{bm}
\usepackage{amsmath,amssymb}
\begin{document}
\title{Simulation of fault-tolerant quantum circuits on
quantum computational tensor network}  
\author{Tomoyuki Morimae}
\affiliation{
LAMA,
Universit\'e Paris-Est Marne-la-Vall\'ee, 77454 Marne-la-Vall\'ee
Cedex 2, France
}
\affiliation{
IRCS,
Tokyo Institute of Technology, 2-12-1 Ookayama, Meguro-ku,
Tokyo 152-8550, Japan
}

\author{Keisuke Fujii}
\affiliation{Graduate School of Engineering Science,
Osaka University, Toyonaka, Osaka 560-8531, Japan}
\date{\today}
            
\begin{abstract}
In the framework of quantum computational tensor network [D. Gross and J. Eisert, Phys. Rev. Lett. {\bf98}, 220503 (2007)], which is
a general framework of measurement-based quantum computation, 
the resource many-body state
is represented in a tensor-network form (or a matrix-product form), 
and universal quantum computation
is performed in a virtual linear space, which is called a
correlation space, where tensors live.
Since any unitary operation, state preparation, and the projection
measurement in the computational basis can be simulated
in a correlation
space, it is natural to expect that fault-tolerant
quantum circuits
can also be simulated in a correlation space.
However, we point out that 
not all physical errors  
on physical qudits 
appear as linear completely-positive trace-preserving errors 
in a correlation space.
Since
the theories of fault-tolerant quantum circuits known so far
assume such noises,
this means that
the simulation of fault-tolerant quantum circuits in
a correlation space is not so straightforward for
general resource states.

\end{abstract}
\pacs{03. 67. -a}
\maketitle  

\section{Introduction}
Quantum many-body states, which have long been central research
objects in condensed matter physics, statistical physics,
and quantum chemistry,
are now attracting the renewed interest in quantum information science
as fundamental resources for quantum information processing.
One of the most celebrated examples
is one-way 
quantum computation~\cite{one-way,one-way2,one-way3}. 
Once the highly-entangled
many-body state which is called the cluster state
is prepared, universal quantum computation is possible with
adaptive local measurements on each qubit.
Recently, the concept of 
quantum computational tensor network (QCTN)~\cite{Gross1,Gross2,Gross3},
which is the general framework of measurement-based
quantum computation on quantum many-body states,
was proposed. 
This novel framework has enabled us to understand how 
general measurement-based quantum computation is 
performed on many other resource states
beyond the cluster state.
The most innovative feature of QCTN is that the resource
state is represented in   
a tensor-network form 
(or a matrix-product form)~\cite{Fannes,Verstraete,Cirac}, 
and universal quantum computation
is performed in the virtual linear space where tensors live.
For example, let us consider the one-dimensional open-boundary 
chain of $N$ qudits 
in the matrix-product form
\begin{eqnarray}
|\Psi(L,R)\rangle_{1}^N\equiv\frac{1}{\sqrt{f_N(|L\rangle,|R\rangle)}}
\sum_{k_1=0}^{d-1}...\sum_{k_N=0}^{d-1}
\langle L|A[k_N]...A[k_1]|R\rangle
|k_N,...,k_1\rangle,
\label{MPS}
\end{eqnarray}
where 
\begin{eqnarray*}
f_N(|L\rangle,|R\rangle)\equiv\langle L|({\mathcal A}^N|R\rangle\langle R|)|L\rangle
\end{eqnarray*}
is the normalization factor,
\begin{eqnarray*}
{\mathcal A}\rho\equiv\sum_{i=0}^{d-1}A[i]\rho A^\dagger[i]
\end{eqnarray*}
is a map,
$\{|0\rangle,...,|d-1\rangle\}$ is a certain basis in 
the $d$-dimensional
Hilbert space ($2\le d<\infty$),
$|L\rangle$ and $|R\rangle$ are $D$-dimensional complex
vectors, and $\{A[0],...,A[d-1]\}$ are $D\times D$ complex matrices.
Let us also define the projection measurement $\mathcal{M}_{\theta,\phi}$ 
on a single physical qudit by
\begin{eqnarray}
{\mathcal M}_{\theta,\phi}\equiv\big\{|\alpha_{\theta,\phi}\rangle,
|\beta_{\theta,\phi}\rangle,
|2\rangle,...,|d-1\rangle \big\},
\label{measurement}
\end{eqnarray}
where
\begin{eqnarray*}
|\alpha_{\theta,\phi}\rangle
&\equiv&\cos\frac{\theta}{2}|0\rangle+e^{i\phi}\sin\frac{\theta}{2}|1\rangle,\\
|\beta_{\theta,\phi}\rangle
&\equiv&\sin\frac{\theta}{2}|0\rangle-e^{i\phi}\cos\frac{\theta}{2}|1\rangle,
\end{eqnarray*}
$0<\theta<\pi$, and $0\le \phi < 2\pi$.
If we do the measurement ${\mathcal M}_{\theta,\phi}$ on the first 
physical qudit 
of $|\Psi(L,R)\rangle_1^N$ and if
the first physical qudit is projected onto, for example, 
$|\alpha_{\theta,\phi}\rangle$
as a result of this measurement,
the state $|\Psi(L,R)\rangle_1^N$ becomes
\begin{eqnarray*}
&&
\frac{1}{\sqrt{f_{N-1}(|L\rangle,A[\alpha_{\theta,\phi}]|R\rangle)}}
\sum_{k_2=0}^{d-1}
...
\sum_{k_N=0}^{d-1}
\langle L|A[k_N]...A[k_2]A[\alpha_{\theta,\phi}]
|R\rangle
|k_N,...,k_2\rangle\otimes|\alpha_{\theta,\phi}\rangle\\
&=&
\frac{1}{\sqrt{f_{N-1}(|L\rangle,A[\alpha_{\theta,\phi}]|R\rangle)}}
\|A[\alpha_{\theta,\phi}]\|
\sum_{k_2=0}^{d-1}
...
\sum_{k_N=0}^{d-1}
\langle L|A[k_N]...A[k_2]\frac{A[\alpha_{\theta,\phi}]}
{\|A[\alpha_{\theta,\phi}]\|}
|R\rangle
|k_N,...,k_2\rangle\otimes|\alpha_{\theta,\phi}\rangle,
\end{eqnarray*}
where 
\begin{eqnarray*}
A[\alpha_{\theta,\phi}]\equiv\cos\frac{\theta}{2}A[0]+
e^{-i\phi}\sin\frac{\theta}{2}A[1].
\end{eqnarray*}
Then, we say ``the operation 
\begin{eqnarray*}
|R\rangle\to
\frac{A[\alpha_{\theta,\phi}]}{\|A[\alpha_{\theta,\phi}]\|}
|R\rangle
\end{eqnarray*}
is implemented in the correlation space".
In particular,
if $A[0]$, $A[1]$, $\theta$, and $\phi$
are appropriately chosen in such a way that 
$A[\alpha_{\theta,\phi}]$ is proportional to a unitary,
we can ``simulate" the unitary evolution 
\begin{eqnarray*}
\frac{A[\alpha_{\theta,\phi}]}{\|A[\alpha_{\theta,\phi}]\|}|R\rangle
\end{eqnarray*}
of the vector $|R\rangle$ in
the virtual linear space where $A$'s, $|R\rangle$, and $|L\rangle$ live.
This virtual
linear space is called the correlation space~\cite{Gross1,Gross2,Gross3,
Download,Upload,FM}.
The core of 
QCTN 
is this ``virtual quantum computation" in the correlation space. 
If the correlation space has a sufficient structure
and if $A$'s, $|L\rangle$, and $|R\rangle$ are appropriately chosen,
we can ``simulate" universal quantum circuit in the correlation 
space~\cite{Gross1,Gross2,Gross3,Download,Upload,FM}.

For the realization of a scalable quantum computer, 
a theory of fault-tolerant (FT)
quantum computation~\cite{Shor,Aharonov,Kitaev,Knill,Terhal} is necessary.
In fact, several researches have been performed
on FT quantum computation
in the one-way model~\cite{one-way3,ND05,AL06,Silva,Fujii1,Fujii2}.
However, 
there has been no result about a theory of FT quantum computation 
on general QCTN~\cite{recently}.
In particular, there is severe lack of knowledge about
FT quantum computation
on resource states with $d\ge3$.
It is necessary to  
consider 
resource states with $d\ge3$
if we want to enjoy the cooling preparation
of a resource state
and the energy-gap protection of measurement-based quantum computation
with a physically natural Hamiltonian, 
since no genuinely entangled qubit state can be
the unique ground state of a two-body frustration-free
Hamiltonian~\cite{no-go}. 

One straightforward way of implementing FT quantum computation
on QCTN
is to encode physical qudits with a quantum error correcting
code:
\begin{eqnarray*}
|\tilde{\Psi}\rangle\equiv
\frac{1}{\sqrt{f_N(L,R)}}
\sum_{k_1=0}^{d-1}...\sum_{k_N=0}^{d-1}
\langle L|A[k_N]...A[k_1]|R\rangle
|\tilde{k}_N,...,\tilde{k}_1\rangle,
\end{eqnarray*}
where $|\tilde{k}_i\rangle$ ($i=1,...,N$) is the encoded version
of $|k_i\rangle$ (such as $|\tilde{0}\rangle=|000\rangle$ and $|\tilde{1}\rangle=|111\rangle$, etc.)
In fact, this strategy was taken in Refs.~\cite{Fujii1,Fujii2}
for the one-way model $(d=2)$,  
and it was shown there that a FT construction of 
the encoded cluster state
is possible.
For $d\ge3$, however,
such a strategy is difficult, since
theories of  
quantum error correcting codes 
and 
FT preparations of 
the encoded resource 
state $|\tilde{\Psi}\rangle$ 
are less developed for $d\ge3$.
Furthermore, if we encode physical qudits with a quantum error
correcting code, the parent
Hamiltonian should no longer be two-body interacting one.

The other way of implementing FT quantum computation
on QCTN
is to simulate FT quantum circuits
in the correlation space.
Since any unitary operation, state preparation,
and the projective measurement 
in the computational basis
can be simulated in a correlation space (for a more precise discussion
about the possibility of the measurement, see Ref.~\cite{FM}),  
it is natural to expect that FT quantum circuits
can also be simulated in a correlation space.
An advantage of this strategy is that 
theories of FT quantum circuits
for qubit systems are well developed~\cite{Shor,Aharonov,Kitaev,Knill,Terhal}. 
In fact, this strategy was taken in Refs.~\cite{ND05,AL06} 
for the one-way model $(d=2)$.
They introduced a method (which we call ``the ensemble method"
since the ensemble of all measurement results are considered)
of simulating quantum circuits in the correlation space
of the cluster state, and showed that all physical errors
on physical qubits can be linear completely-positive
trace-preserving (CPTP) maps in the correlation space of
the cluster state.
This means that FT quantum circuits can be simulated in the correlation
space of the cluster state.

In this paper, however,
we point out that it is not so straightforward to simulate
FT quantum circuits  
in a correlation space of a general resource state.
In the next section, Sec.~\ref{sec:cluster}, 
we review the simulation of FT quantum circuits 
on the one-dimensional cluster state~\cite{ND05,AL06} 
in terms of the QCTN picture to fix the notation.
We see that for the cluster state all physical errors can be
linear completely-positive trace-preserving (CPTP) 
maps in the correlation space, and therefore 
the theory of FT quantum circuits
can be used in the correlation space.
However, this is not the case for other general resource states
of QCTN.
As an example, we consider a similar way of simulating 
quantum circuits
in the correlation space
of the one-dimensional AKLT state~\cite{Brennen,AKLT}
in Sec.~\ref{sec:AKLT},
and show that not all physical errors can be linear CPTP maps
in the correlation space of the AKLT state.
Since all theories of FT quantum circuits
known so far
assume such noises~\cite{Shor,Aharonov,Kitaev,Knill,Terhal},
this means that it is not so straightforward to apply these FT theories
to quantum circuits simulated in the correlation space
of general QCTN.
In Sec.~\ref{sec:why},
we give some intuitive explanations of the reason why
the cluster state is so special, and why not all resource states
work as the cluster state.
In Sec.~\ref{sec:pure},
we consider another standard way of simulating quantum
circuits in the correlation space,
which we call ``the trajectory method" since a specific trajectory
(measurement results) is considered.
However, we show a general theorem that such
an another way does neither work
if
$d\ge3$.

In short, we show in this paper that it is not so straightforward to 
simulate
FT quantum circuits in the correlation space of a general
resource state.
Since all errors behave nicely in the correlation space of
the cluster state~\cite{ND05,AL06},
less attention has been paid to the difference between
a real physical space and a correlation space of a general
resource state.
Our results here suggest that
these two spaces can be different, and because of the difference,
simulations of FT quantum circuits can be difficult in a
correlation space.
Of course, we do not show here the impossibility of making a QCTN fault-tolerant.
In a future, a highly elaborated method might be found
which makes all QCTN fault-tolerant.
We hope that our results will help to study such a challenging subject
of a future study.

{\bf Assumptions}: Throughout this paper, we make the following
assumptions: Since the MPS $|\Psi(L,R)\rangle_1^N$, Eq.~(\ref{MPS}), is a resource state for measurement-based
quantum computation,
we can assume without loss of generality that 
$A[\alpha_{\theta,\phi}]$, $A[\beta_{\theta,\phi}]$, $A[2]$, 
$A[3]$, ..., $A[d-1]$
are unitary
up to constants:
\begin{eqnarray}
A[\alpha_{\theta,\phi}]&=&c_\alpha U_\alpha,\nonumber\\
A[\beta_{\theta,\phi}]&=&c_\beta U_\beta,\nonumber\\
A[2]&=&c_2 U_2,\nonumber\\
A[3]&=&c_3 U_3,\nonumber\\
&...&\nonumber\\
A[d-1]&=&c_{d-1} U_{d-1},
\label{assumption0}
\end{eqnarray}
where $c_\alpha$, $c_\beta$, $c_2$, ... $c_{d-1}$ are real positive numbers,
$U_\alpha$, $U_\beta$, $U_2$, ..., $U_{d-1}$ are unitary operators,
and
\begin{eqnarray*}
A[\beta_{\theta,\phi}]\equiv\sin\frac{\theta}{2}A[0]-e^{-i\phi}\cos\frac{\theta}{2}A[1].
\end{eqnarray*}
This means that
any operation implemented in the correlation space
by the measurement ${\mathcal M}_{\theta,\phi}$ on a single
physical qudit of $|\Psi(L,R)\rangle_1^N$ is unitary.
Note that this assumption is reasonable, since otherwise 
$|\Psi(L,R)\rangle_1^N$ does not seem to be useful as a resource for measurement-based quantum computation.
In fact, all known resource states so far~\cite{one-way,one-way2,one-way3,Gross1,Gross2,Gross3,Brennen,Caimagnet,tricluster,Miyake2dAKLT},
including the cluster state and the AKLT state, 
satisfy this assumption by appropriately rotating each local physical basis.
Furthermore, we can take $c_\alpha$, $c_\beta$, $c_2$,..., $c_{d-1}$ such that 
\begin{eqnarray*}
C\equiv c_\alpha^2+c_\beta^2+\sum_{k=2}^{d-1}c_k^2=1,
\end{eqnarray*}
since
\begin{eqnarray*}
\frac{1}{\sqrt{f_N(|L\rangle,|R\rangle)}}
\sum_{k_1,...,k_N}\langle L|A[k_N]...A[k_1]|R\rangle
|k_N,...,k_1\rangle
=
\frac{1}{\sqrt{f_N(|L\rangle,|R\rangle)}}\sqrt{C}^N
\sum_{k_1,...,k_N}\langle L|\frac{A[k_N]}{\sqrt{C}}...\frac{A[k_1]}{\sqrt{C}}|R\rangle
|k_N,...,k_1\rangle
\end{eqnarray*}
and we can redefine $A[k_i]/\sqrt{C}\to A[k_i]$.

\section{Simulation on the cluster state}
\label{sec:cluster}
Let us first review the results for the cluster state~\cite{ND05,AL06}
in terms of the QCTN picture to fix the notation.

\subsection{Simulation on the cluster state without error}
Let us first assume that there is no error.
The one-dimensional cluster state is the matrix-product state
defined by
$d=2$, $A[0]=|+\rangle\langle0|$, 
and $A[1]=|-\rangle\langle1|$.
We measure each physical qubit in the basis 
\begin{eqnarray*}
|\theta_s\rangle&\equiv&\frac{1}{\sqrt{2}}(
|0\rangle+(-1)^se^{i\theta}|1\rangle),
\end{eqnarray*}
where $s\in\{0,1\}$.
In the correlation space, 
$X^sJ(\theta)$,
where
$J(\theta)\equiv He^{i\theta Z/2}$,
is implemented.
First, we measure the first physical qubit in
the $\{|\theta_0\rangle,|\theta_1\rangle\}$ basis.
Then we obtain
\begin{eqnarray*}
\frac{1}{2f_N(|L\rangle,|R\rangle)}
\sum_{s_1=0}^1
W(X^{s_1}J(\theta)|R\rangle)_2\otimes 
|\theta_{s_1}\rangle\langle\theta_{s_1}|
\otimes
m(s_1),
\end{eqnarray*}
where
$m(0)$ and $m(1)$ are mutually orthogonal states
which record the measurement result,
and
\begin{eqnarray*}
W(|\psi\rangle)_r\equiv
\sum_{k_r,...,k_N}\sum_{k_r',...,k_N'}
\langle L|A[k_N]...A[k_r]|\psi\rangle\langle\psi|
A^\dagger[k_r']...
A^\dagger[k_N']|L\rangle
|k_N,...,k_r\rangle\langle k_N',...,k_r'|.
\end{eqnarray*}

If we trace out the measured first physical qubit 
$|\theta_{s_1}\rangle\langle\theta_{s_1}|$, 
we obtain
\begin{eqnarray*}
\frac{1}{2f_N(|L\rangle,|R\rangle)}
\sum_{s_1=0}^1
W(X^{s_1}J(\theta)|R\rangle)_2\otimes 
m(s_1).
\end{eqnarray*}

Second, we measure 
the second physical qubit
in the 
$\{X^{s_1}|\phi_0\rangle,X^{s_1}|\phi_1\rangle\}$
basis.
Then we obtain
\begin{eqnarray*}
&&
\frac{1}{2^2f_N(|L\rangle,|R\rangle)}
\sum_{s_2=0}^1
\sum_{s_1=0}^1
W(X^{s_2}J((-1)^{s_1}\phi)X^{s_1}J(\theta)|R\rangle)_3\otimes
X^{s_1}|\phi_{s_2}\rangle\langle\phi_{s_2}|X^{s_1}
\otimes m(s_1)
\otimes
m(s_2)\\
&=&
\frac{1}{2^2f_N(|L\rangle,|R\rangle)}
\sum_{s_2=0}^1
\sum_{s_1=0}^1
W(X^{s_2}Z^{s_1}J(\phi)J(\theta)|R\rangle)_3\otimes
X^{s_1}|\phi_{s_2}\rangle\langle\phi_{s_2}|X^{s_1}
\otimes
m(s_1)
\otimes
m(s_2).
\end{eqnarray*}
If we trace out the measured second physical qubit
$X^{s_1}|\phi_{s_2}\rangle\langle\phi_{s_2}|X^{s_1}$,
we obtain
\begin{eqnarray*}
\frac{1}{2^2f_N(|L\rangle,|R\rangle)}
\sum_{s_2=0}^1
\sum_{s_1=0}^1
W(X^{s_2}Z^{s_1}J(\phi)J(\theta)|R\rangle)_3\otimes
m(s_1)
\otimes
m(s_2).
\end{eqnarray*}

Third, we measure the third physical qubit
in the $\{Z^{s_1}X^{s_2}|\eta_0\rangle,Z^{s_1}X^{s_2}|\eta_1\rangle\}$
basis.
Then we obtain
\begin{eqnarray*}
&&
\frac{1}{2^3f_N(|L\rangle,|R\rangle)}
\sum_{s_3=0}^1
\sum_{s_2=0}^1
\sum_{s_1=0}^1
W(X^{s_3}X^{s_1}
J((-1)^{s_2}\eta)X^{s_2}Z^{s_1}J(\phi)J(\theta)|R\rangle)_4
\otimes
Z^{s_1}X^{s_2}|\eta_{s_3}\rangle\langle\eta_{s_3}|X^{s_2}Z^{s_1}\\
&&\otimes
m(s_1)
\otimes
m(s_2)
\otimes
m(s_3)
\\
&=&
\frac{1}{2^3f_N(|L\rangle,|R\rangle)}
\sum_{s_3=0}^1
\sum_{s_2=0}^1
\sum_{s_1=0}^1
W(X^{s_3}
Z^{s_2}
J(\eta)J(\phi)J(\theta)|R\rangle)_4
\otimes
Z^{s_1}X^{s_2}|\eta_{s_3}\rangle\langle\eta_{s_3}|X^{s_2}Z^{s_1}
\\
&&
\otimes
m(s_1)
\otimes
m(s_2)
\otimes
m(s_3).
\end{eqnarray*}
If we trace out the measured third physical qubit
$Z^{s_1}X^{s_2}|\eta_{s_3}\rangle\langle\eta_{s_3}|X^{s_2}Z^{s_1}$,
we obtain
\begin{eqnarray*}
&&
\frac{1}{2^3f_N(|L\rangle,|R\rangle)}
\sum_{s_3=0}^1
\sum_{s_2=0}^1
\sum_{s_1=0}^1
W(X^{s_3}
Z^{s_2}
J(\eta)J(\phi)J(\theta)|R\rangle)_4
\otimes
m(s_1)
\otimes
m(s_2)
\otimes
m(s_3)\\
&=&
\frac{1}{2^3f_N(|L\rangle,|R\rangle)}
\sum_{s_3=0}^1
\sum_{s_2=0}^1
W(X^{s_3}
Z^{s_2}
J(\eta)J(\phi)J(\theta)|R\rangle)_4
\otimes
I
\otimes
m(s_2)
\otimes
m(s_3).
\end{eqnarray*}
In this way, we can simulate
the desired unitary operation $J(\eta)J(\phi)J(\theta)$
on $|R\rangle$
up to Pauli byproducts $X^{s_3}Z^{s_2}$ 
in the correlation space. These Pauli byproducts
can be corrected later, since they are specified by
$m(s_2)\otimes m(s_3)$.

\subsection{Effect of a CPTP error on a physical qudit of a general state}
Before studying the simulation of quantum circuits
on the cluster state with error,
let us consider the effect of a CPTP error on a physical qudit
of general resource states,
since we will use it later.
Let us assume that a CPTP error 
\begin{eqnarray*}
\rho\to\sum_{j=1}^wF_j\rho F_j^\dagger,
\end{eqnarray*}
where $\sum_{j=1}^wF_j^\dagger F_j=I$,
occurs on the first physical
qudit of $|\Psi(L,R)\rangle_1^N$:
\begin{eqnarray*}
\frac{1}{f_N(|L\rangle,|R\rangle)}
\sum_j
\sum_{k_1,...,k_N}
\sum_{k_1',...,k_N'}
\langle L|A[k_N]...A[k_1]|R\rangle
\langle R|A^\dagger[k_1']...A^\dagger[k_N']|L\rangle
|k_N,...,k_2\rangle\langle k_2',...,k_N'|\otimes
F_j|k_1\rangle\langle k_1'|F^\dagger_j.
\end{eqnarray*}
If we measure the first physical qudit in a certain basis $\{|m_s\rangle\}$,
\begin{eqnarray*}
&&
\frac{1}{f_N(|L\rangle,|R\rangle)}
\sum_{j,s}
\sum_{k_1,...,k_N}
\sum_{k_1',...,k_N'}
\langle L|A[k_N]...A[k_1]|R\rangle
\langle R|A^\dagger[k_1']...A^\dagger[k_N']|L\rangle
|k_N,...,k_2\rangle\langle k_2',...,k_N'|\\
&&\otimes
|m_s\rangle\langle m_s|
F_j|k_1\rangle\langle k_1'|F^\dagger_j
|m_s\rangle\langle m_s|\\
&=&
\frac{1}{f_N(|L\rangle,|R\rangle)}
\sum_{j,s}
\sum_{k_2,...,k_N}
\sum_{k_2',...,k_N'}
\langle L|A[k_N]...A[k_2]E_{j,s}|R\rangle
\langle R|E^\dagger_{j,s}A^\dagger[k_2']...A^\dagger[k_N']|L\rangle
|k_N,...,k_2\rangle\langle k_2',...,k_N'|\otimes
|m_s\rangle\langle m_s|\\
&=&
\frac{1}{f_N(|L\rangle,|R\rangle)}
\sum_s
\Big(
\sum_j
W(
E_{j,s}|R\rangle)_2
\Big)
\otimes
|m_s\rangle\langle m_s|,
\end{eqnarray*}
where
\begin{eqnarray*}
E_{j,s}\equiv
\sum_{k}A[k]
\langle m_s|F_j|k\rangle.
\end{eqnarray*}
If we trace out $|m_s\rangle$,
we obtain
\begin{eqnarray*}
&&
\frac{1}{f_N(|L\rangle,|R\rangle)}\sum_s\sum_jW(E_{j,s}|R\rangle)_2\\
&&=\frac{1}{f_N(|L\rangle,|R\rangle)}
\sum_{k_2,...,k_N}
\sum_{k_2',...,k_N'}
\langle L|A[k_N]...A[k_2]
\Big(\sum_{j,s}E_{j,s}|R\rangle
\langle R|E^\dagger_{j,s}\Big)A^\dagger[k_2']...A^\dagger[k_N']|L\rangle
|k_N,...,k_2\rangle\langle k_2',...,k_N'|.
\end{eqnarray*}
This means that the map
\begin{eqnarray}
|R\rangle\langle R|\to
\sum_{s,j}E_{j,s}|R\rangle\langle R|E^\dagger_{j,s}
\label{map}
\end{eqnarray}
is implemented in the correlation space.
Note that
\begin{eqnarray*}
\sum_{j,s}E^\dagger_{j,s}
E_{j,s}&=&\sum_{j,s,k,k'}
A^\dagger[k]A[k']\langle k|F^\dagger_j|m_s\rangle
\langle m_s|F_j|k'\rangle\\
&=&
\sum_{k,k'}
A^\dagger[k]A[k']\langle k|
k'\rangle\\
&=&
\sum_{k,k'}
A^\dagger[k]A[k']\langle k|
U_{\mathcal M_{\theta,\phi}} U_{\mathcal M_{\theta,\phi}}^\dagger
|k'\rangle\\
&=&
A^\dagger[\alpha_{\theta,\phi}]A[\alpha_{\theta,\phi}]
+A^\dagger[\beta_{\theta,\phi}]A[\beta_{\theta,\phi}]
+\sum_{k=2}^{d-1}
A^\dagger[k]A[k]\\
&=&I,
\end{eqnarray*}
where 
\begin{eqnarray*}
U_{\mathcal M_{\theta,\phi}}\equiv
|\alpha_{\theta,\phi}\rangle\langle0|
+|\beta_{\theta,\phi}\rangle\langle1|
+\sum_{k=2}^{d-1}|k\rangle\langle k|
\end{eqnarray*}
is a unitary operator.
Therefore, the map Eq.~(\ref{map}) is CPTP.

Note that this result does not mean that we can always have CPTP errors
in the correlation space: In this section, we did not consider
any quantum gate. If we implement quantum gates in the correlation
space, the situation becomes more complicated, and,
as we will see later, we sometimes have non-CPTP errors in
the correlation space.

\subsection{Simulation on the cluster state with error}
Now let us consider the case where we implement quantum gates
on the cluster state with error.
We assume that
a CPTP error occurs on the first physical qubit
of the one-dimensional cluster state.
If we measure the first physical qubit in the
$\{|\theta_0\rangle,|\theta_1\rangle\}$ basis
after such an error,
we obtain
\begin{eqnarray*}
\frac{1}{f_N(|L\rangle,|R\rangle)}
\sum_{s_1=0}^1
\Big(\sum_j
W(E_{j,s_1}|R\rangle)_2
\Big)
\otimes 
|\theta_{s_1}\rangle\langle\theta_{s_1}|
\otimes
m(s_1).
\end{eqnarray*}
By tracing out the first physical qubit 
$|\theta_{s_1}\rangle\langle\theta_{s_1}|$,
we obtain
\begin{eqnarray*}
\frac{1}{f_N(|L\rangle,|R\rangle)}
\sum_{s_1=0}^1
\Big(\sum_j
W(E_{j,s_1}|R\rangle)_2
\Big)
\otimes 
m(s_1).
\end{eqnarray*}

Second, we measure 
the second physical qubit
in the 
$\{X^{s_1}|\phi_0\rangle,X^{s_1}|\phi_1\rangle\}$
basis.
Then we obtain
\begin{eqnarray*}
\frac{1}{2f_N(|L\rangle,|R\rangle)}
\sum_{s_2=0}^1
\sum_{s_1=0}^1
\Big(
\sum_j
W(X^{s_2}J((-1)^{s_1}\phi)E_{j,s_1}|R\rangle)_3\Big)\otimes
X^{s_1}|\phi_{s_2}\rangle\langle\phi_{s_2}|X^{s_1}
\otimes
m(s_1)
\otimes m(s_2).
\end{eqnarray*}
By tracing out the second physical qubit
$X^{s_1}|\phi_{s_2}\rangle\langle\phi_{s_2}|X^{s_1}$,
\begin{eqnarray*}
\frac{1}{2f_N(|L\rangle,|R\rangle)}
\sum_{s_2=0}^1
\sum_{s_1=0}^1
\Big(\sum_j
W(X^{s_2}J((-1)^{s_1}\phi)E_{j,s_1}|R\rangle)_3\Big)\otimes
m(s_1)
\otimes m(s_2).
\end{eqnarray*}

Third, we measure the third physical qubit
in the $\{Z^{s_1}X^{s_2}|\eta_0\rangle,Z^{s_1}X^{s_2}|\eta_1\rangle\}$
basis.
Then we obtain
\begin{eqnarray*}
&&
\frac{1}{2^2f_N(|L\rangle,|R\rangle)}
\sum_{s_3,s_2,s_1}
\Big(\sum_j
W(X^{s_3}X^{s_1}J((-1)^{s_2}\eta)
X^{s_2}J((-1)^{s_1}\phi)E_{j,s_1}|R\rangle)_4
\Big)
\otimes
Z^{s_1}X^{s_2}|\eta_{s_3}\rangle\langle\eta_{s_3}|X^{s_2}Z^{s_1}\\
&&\otimes m(s_1)
\otimes m(s_2)
\otimes m(s_3)\\
&=&
\frac{1}{2^2f_N(|L\rangle,|R\rangle)}
\sum_{s_3,s_2,s_1}
\Big(
\sum_j
W(X^{s_3}X^{s_1}Z^{s_2}J(\eta)
J((-1)^{s_1}\phi)E_{j,s_1}|R\rangle)_4
\Big)
\otimes
Z^{s_1}X^{s_2}|\eta_{s_3}\rangle\langle\eta_{s_3}|X^{s_2}Z^{s_1}\\
&&\otimes m(s_1)
\otimes m(s_2)
\otimes m(s_3).
\end{eqnarray*}
If we trace out the third physical qubit
$Z^{s_1}X^{s_2}|\eta_{s_3}\rangle\langle\eta_{s_3}|X^{s_2}Z^{s_1}$,
\begin{eqnarray*}
\frac{1}{2^2f_N(|L\rangle,|R\rangle)}
\sum_{s_3,s_2,s_1}
\Big(\sum_j
W(X^{s_3}X^{s_1}Z^{s_2}J(\eta)
J((-1)^{s_1}\phi)E_{j,s_1}|R\rangle)_4
\Big)
\otimes m(s_1)
\otimes m(s_2)
\otimes m(s_3).
\end{eqnarray*}
If we further trace out the first record state $m(s_1)$,
\begin{eqnarray*}
&&
\frac{1}{2^2f_N(|L\rangle,|R\rangle)}
\sum_{s_3,s_2,j}
\Big(
W(X^{s_3}Z^{s_2}J(\eta)
J(\phi)E_{j,0}|R\rangle)_4
+
W(X^{s_3}XZ^{s_2}J(\eta)
J(-\phi)E_{j,1}|R\rangle)_4
\Big)
\otimes m(s_2)
\otimes m(s_3)\\
&=&
\frac{1}{2^2f_N(|L\rangle,|R\rangle)}
\sum_{s_3,s_2,j}
\Big(
W(X^{s_3}Z^{s_2}J(\eta)
J(\phi)E_{j,0}|R\rangle)_4
+
W(X^{s_3}Z^{s_2}J(\eta)
J(\phi)XE_{j,1}|R\rangle)_4
\Big)
\otimes m(s_2)
\otimes m(s_3).
\end{eqnarray*}
In other words, the map
\begin{eqnarray*}
|R\rangle\langle R|\to
\sum_j\Big(
E_{j,0}|R\rangle\langle R|E^\dagger_{j,0}
+
XE_{j,1}|R\rangle\langle R|E^\dagger_{j,1}X
\Big)
\end{eqnarray*}
is implemented in the correlation space
up to the rotation $J(\eta)J(\phi)$
and a Pauli byproduct $X^{s_3}Z^{s_2}$.

Since
\begin{eqnarray*}
\sum_j\Big(
E^\dagger_{j,0}E_{j,0}
+
(E^\dagger_{j,1}X)(XE_{j,1})
\Big)
=
\sum_{j,s}E^\dagger_{j,s}E_{j,s}=I,
\end{eqnarray*}
this is CPTP error.
In short, a CPTP error on a physical qubit
becomes a CPTP error in the correlation space of the cluster state.
As is shown in Appendices~\ref{App:tri1} and \ref{App:tri2},
similar result is obtained for the tricluster state~\cite{tricluster},
which is a variant of the cluster state.

\section{AKLT state}
\label{sec:AKLT}
We have seen in the previous section
that CPTP errors on physical qubits of the one-dimensional cluster
state become linear CPTP maps in the correlation space.
However, this is not always the case for general resource states.
In order to see it, let us consider the one-dimensional AKLT state
as an example.

\subsection{Simulation on the AKLT state without error}
First we assume there is no error.
The one-dimensional AKLT state
is the matrix-product state defined by
$d=3$,
\begin{eqnarray*}
A[0]&=&\frac{1}{\sqrt{3}}X,\\
A[1]&=&\frac{1}{\sqrt{3}}XZ,\\
A[2]&=&\frac{1}{\sqrt{3}}Z.
\end{eqnarray*}


If we measure the first physical qutrit
of the AKLT state
in the basis
\begin{eqnarray*}
{\mathcal M}_{\theta,\pi/2}\equiv\Big\{
\cos\frac{\theta}{2}|0\rangle
+i\sin\frac{\theta}{2}|1\rangle,
\sin\frac{\theta}{2}|0\rangle
-i\cos\frac{\theta}{2}|1\rangle,
|2\rangle\Big\},
\end{eqnarray*}
we obtain
\begin{eqnarray*}
\frac{1}{3f_N(|L\rangle,|R\rangle)}
\sum_{s_1=0}^2
W(Q_1(s_1)|R\rangle)_2\otimes
\rho_1(s_1)\otimes m(s_1),
\end{eqnarray*}
where $\rho_1(s_1)$ is the state of the first physical qutrit
after the measurement,
$m(s_1)$ is the register state which records the first measurement
result,
and
\begin{eqnarray*}
Q_1(0)&=&XS_Z(\theta),\\
Q_1(1)&=&XZS_Z(\theta),\\
Q_1(2)&=&Z.
\end{eqnarray*}
Here, $S_Z(\theta)\equiv e^{-iZ\theta/2}$.
By tracing out the first measured physical qutrit
$\rho_1(s_1)$,
we obtain
\begin{eqnarray*}
\frac{1}{3f_N(|L\rangle,|R\rangle)}
\sum_{s_1=0}^2
W(Q_1(s_1)|R\rangle)_2\otimes
m(s_1).
\end{eqnarray*}

Next we measure the second physical qutrit
by choosing measurement basis according to $s_1$~\cite{Brennen}.
Then,
we obtain
\begin{eqnarray*}
\frac{1}{3^2f_N(|L\rangle,|R\rangle)}
\sum_{s_2=0}^2
\sum_{s_1=0}^2
W(Q_2(s_1,s_2)Q_1(s_1)|R\rangle)_3\otimes
\rho_2(s_2,s_1)
\otimes
m(s_1)
\otimes
m(s_2),
\end{eqnarray*}
where
$\rho_2(s_1,s_2)$ is the state of the second physical qutrit
after the measurement,
$m(s_2)$ is the register state which records the second measurement result,
and
\begin{eqnarray*}
Q_2(s_1,0)&=&XS_Z(\theta),\\
Q_2(s_1,1)&=&XZS_Z(\theta),\\
Q_2(s_1,2)&=&Z,\\
\end{eqnarray*}
if $s_1=2$
and
\begin{eqnarray*}
Q_2(s_1,0)&=&X,\\
Q_2(s_1,1)&=&XZ,\\
Q_2(s_1,2)&=&Z,
\end{eqnarray*}
for other $s_1$.
By tracing out the measured second physical qutrit
$\rho_2(s_2,s_1)$,
\begin{eqnarray*}
\frac{1}{3^2f_N(|L\rangle,|R\rangle)}
\sum_{s_2=0}^2
\sum_{s_1=0}^2
W(Q_2(s_1,s_2)Q_1(s_1)|R\rangle)_3\otimes
m(s_1)
\otimes
m(s_2).
\end{eqnarray*}

If we repeat these process, after measuring the $r$th physical qutrit,
we obtain
\begin{eqnarray*}
\frac{1}{3^rf_N(|L\rangle,|R\rangle)}
\sum_{s_1=0}^2...
\sum_{s_r=0}^2
W(Q_r(s_1,...,s_r)...Q_2(s_1,s_2)Q_1(s_1)|R\rangle)_{r+1}
\otimes
m(s_1)
\otimes
...
\otimes
m(s_r),
\end{eqnarray*}
where 
\begin{eqnarray*}
Q_k(s_1,...,s_{k-1},0)&=&XS_Z(\theta),\\
Q_k(s_1,...,s_{k-1},1)&=&XZS_Z(\theta),\\
Q_k(s_1,...,s_{k-1},2)&=&Z,
\end{eqnarray*}
if $s_1=...=s_{k-1}=2$
and
\begin{eqnarray*}
Q_k(s_1,...,s_{k-1},0)&=&X,\\
Q_k(s_1,...,s_{k-1},1)&=&XZ,\\
Q_k(s_1,...,s_{k-1},2)&=&Z,
\end{eqnarray*}
for other $s_1,...,s_{k-1}$.

If $s_1=...=s_r=2$,
\begin{eqnarray*}
Q_r(s_1,...,s_r)...Q_2(s_1,s_2)Q_1(s_1)
=Z^r.
\end{eqnarray*}
For other $s_1,...,s_r$,
\begin{eqnarray*}
Q_r(s_1,...,s_r)...Q_2(s_1,s_2)Q_1(s_1)
=
X^{f(s_1,...,s_r)}Z^{g(s_1,...,s_r)}S_Z(\theta),
\end{eqnarray*}
where 
\begin{eqnarray*}
f(s_1,...,s_r)&=&\bigoplus_{i=1}^r(\delta_{s_i,0}\oplus\delta_{s_i,1}),\\
g(s_1,...,s_r)&=&\bigoplus_{i=1}^r(\delta_{s_i,1}\oplus\delta_{s_i,2}).
\end{eqnarray*}

Let us add the flag state $\eta$ as
\begin{eqnarray*}
&&
\frac{1}{3^rf_N(|L\rangle,|R\rangle)}
\sum_{s_1,...,s_r}
W(
Q_r(s_1,...,s_r)...Q_2(s_1,s_2)Q_1(s_1)
|R\rangle)_{r+1}
\otimes
m(s_1)
\otimes
...
\otimes
m(s_r)
\otimes
\eta(f(s_1,...,s_r),g(s_1,...,s_r))\\
\end{eqnarray*}
where 
$\eta(0,0)$,
$\eta(0,1)$,
$\eta(1,0)$,
and
$\eta(1,1)$,
are mutually orthogonal with each other.
If we trace out all register states,
$m(s_1)$,
$m(s_2)$,
..., and
$m(s_r)$,
we obtain
\begin{eqnarray}
&&
\frac{1}{3^rf_N(|L\rangle,|R\rangle)}
\sum_{s_1,...,s_r}
W(
Q_r(s_1,...,s_r)...Q_2(s_1,s_2)Q_1(s_1)
|R\rangle)_{r+1}
\otimes
\eta(f(s_1,...,s_r),g(s_1,...,s_r))
\label{z_rotation}\\
&=&
\frac{1}{3^rf_N(|L\rangle,|R\rangle)}
\sum_{p=0}^1
\sum_{q=0}^1
\Big(
\big|S_{p,q}^r\big|
\cdot
W(X^pZ^qS_Z(\theta)|R\rangle)_{r+1}
+h(p,q,r)W(Z^r|R\rangle)_{r+1}
\Big)
\otimes
\eta(p,q)
\label{z_rotation2},
\end{eqnarray}
where 
\begin{eqnarray*}
S_{p,q}^r\equiv\Big\{
(s_1,...,s_r)\in \{0,1,2\}^{\times r}\setminus(2,...,2)~
\Big|~
f(s_1,...,s_r)=p~\mbox{and}~g(s_1,...,s_r)=q
\Big\}
\end{eqnarray*}
and
\begin{eqnarray*}
h(p,q,r)=
\left\{
\begin{array}{cc}
\delta_{p,0}\delta_{q,0}&(r=\mbox{even})\\
\delta_{p,0}\delta_{q,1}&(r=\mbox{odd}).
\end{array}
\right.
\end{eqnarray*}

In this way, we can implement the desired rotation 
$S_Z(\theta)$ 
up to Pauli byproducts $X^pZ^q$. 
Note that in the second term 
\begin{eqnarray*}
h(p,q,r)W(Z^r|R\rangle_{r+1})
\end{eqnarray*}
the desired rotation 
$S_Z(\theta)$ 
is not implemented (the trivial $Z^r$ is implemented, instead). 
However, this term can be treated
as the usual error which can be corrected by the usual
FT circuits.

\subsection{Simulation on the AKLT state with error}
Let us study what happens if an error occurs.
We will see that not all physical errors on a physical 
qutrit can be linear CPTP errors in the correlation space.

If a CPTP error occurs on the first physical qutrit,
Eq.~(\ref{z_rotation})
becomes
\begin{eqnarray*}
&&
\frac{1}{3^{r-1}f_N(|L\rangle,|R\rangle)}
\sum_{s_1,...,s_r}
\sum_j
W(Q_r(s_1,...,s_r)...Q_2(s_1,s_2)E_{j,s_1}|R\rangle)_{r+1}
\otimes
\eta(f(s_1,...,s_r),g(s_1,...,s_r))\\
&=&
\frac{1}{3^{r-1}f_N(|L\rangle,|R\rangle)}
\sum_j
\sum_{p,q}
\Big(
\sum_{(s_1,...,s_r)\in S_{p,q}^r}
W(Q_r(s_1,...,s_r)...Q_2(s_1,s_2)E_{j,s_1}|R\rangle)_{r+1}\\
&&+
h(p,q,r)W(Z^{r-1}E_{j,2}|R\rangle)_{r+1}
\Big)
\otimes
\eta(p,q).
\end{eqnarray*}
This means that
for fixed $p$ and $q$, the map
\begin{eqnarray*}
|R\rangle\langle R|&\to&
\sum_{(s_1,...,s_r)\in S_{p,q}^r}
\sum_j
Q_r(s_1,...,s_r)...Q_2(s_1,s_2)E_{j,s_1}|R\rangle\langle R|
E^\dagger_{j,s_1}
Q_2^\dagger(s_1,s_2)...
Q_r^\dagger(s_1,...,s_r)\\
&&+
h(p,q,r)\sum_j
Z^{r-1}E_{j,2}|R\rangle\langle R|E^\dagger_{j,2}Z^{r-1}
\end{eqnarray*}
is implemented in the correlation space.
Note that
\begin{eqnarray*}
&&
\sum_{(s_1,...,s_r)\in S_{p,q}^r}
\sum_j
E^\dagger_{j,s_1}
Q_2^\dagger(s_1,s_2)...
Q_r^\dagger(s_1,...,s_r)
Q_r(s_1,...,s_r)...Q_2(s_1,s_2)E_{j,s_1}
+h(p,q,r)\sum_j E_{j,2}^\dagger Z^{r-1} Z^{r-1}E_{j,2}\\
&=&
\sum_{(s_1,...,s_r)\in S_{p,q}^r}
\sum_j
E^\dagger_{j,s_1}
E_{j,s_1}
+h(p,q,r)\sum_j E_{j,2}^\dagger E_{j,2}
\\
&=&
\sum_{s_1}
\sum_j
|T_{p,q}^{r,s_1}|\cdot
E^\dagger_{j,s_1}
E_{j,s_1}
+h(p,q,r)\sum_j E_{j,2}^\dagger E_{j,2}
.
\end{eqnarray*}
Here,
\begin{eqnarray*}
T^{r,i}_{p,q}\equiv
\Big\{(s_1,...,s_r)\in\{0,1,2\}^{\times r}\setminus(2,...,2)~\Big|~
s_1=i~\mbox{and}~f(s_1,...,s_r)=p~\mbox{and}~g(s_1,...,s_r)=q\Big\}.
\end{eqnarray*}


For example, let us consider the case $p=1$ and $q=0$.
As is shown in Appendix~\ref{app:T},
\begin{eqnarray*}
|T_{1,0}^{r,0}|-|T_{1,0}^{r,1}|&=&(-1)^{r-1},\\
|T_{1,0}^{r,1}|&=&|T_{1,0}^{r,2}|.
\end{eqnarray*}
Therefore,
\begin{eqnarray*}
\sum_{s_1}\sum_{j=1}^w|T_{1,0}^{r,s_1}|\cdot E_{j,s_1}^\dagger E_{j,s_1}
=
\left\{
\begin{array}{ll}
|T_{1,0}^{r,1}|I
+\sum_jE_{j,0}^\dagger E_{j,0}&(r=\mbox{odd})\\
|T_{1,0}^{r,0}|I
+\sum_jE_{j,1}^\dagger E_{j,1}+\sum_jE_{j,2}^\dagger E_{j,2}&(r=\mbox{even}).
\end{array}
\right.
\end{eqnarray*}
For example, if 
$w=1$
and
\begin{eqnarray*}
F_1=
U_{{\mathcal M}_{\theta,\phi}}
\Big(
\frac{|0\rangle+|1\rangle}{\sqrt{2}}\langle0|
-\frac{|0\rangle-|1\rangle}{\sqrt{2}}\langle1|
+|2\rangle\langle2|\Big),
\end{eqnarray*}
\begin{eqnarray*}
E_{1,0}&=&\sqrt{\frac{2}{3}}|0\rangle\langle1|\\
E_{1,1}&=&\sqrt{\frac{2}{3}}|1\rangle\langle0|\\
E_{1,2}&=&\frac{1}{\sqrt{3}}Z.
\end{eqnarray*}
Then,
\begin{eqnarray}
\sum_{s_1}\sum_{j=1}^w|T_{1,0}^{r,s_1}|\cdot E_{j,s_1}^\dagger E_{j,s_1}
=
\left\{
\begin{array}{ll}
|T_{1,0}^{r,1}|I
+\frac{2}{3}|1\rangle\langle1|&(r=\mbox{odd})\\
|T_{1,0}^{r,0}|I
+\frac{2}{3}|0\rangle\langle0|+\frac{1}{3}I&(r=\mbox{even}),
\end{array}
\right.
\label{nonTPnessAKLTZ}
\end{eqnarray}
which means that the map implemented in the correlation space
is not linear CPTP.

\section{Intuitive explanations}
\label{sec:why}
So far, we have 
seen that the results of Refs.~\cite{ND05,AL06}
for the cluster state
cannot be directly applied to other resource states,
such as the one-dimensional AKLT state.
Why the cluster state is so special?
And why direct applications of the result for the cluster state
to other resource states
do not work?
Although the complete answer to these questions
is beyond the scope of the present paper, since the study 
of QCTN itself has not been fully developed (for example, no one knows 
the necessary and sufficient condition for
tensor-network states to be universal resource states for 
measurement-based quantum computation),
let us try to give some intuitive explanations here.

\begin{figure}[htbp]
\begin{center}
\includegraphics[width=0.8\textwidth]{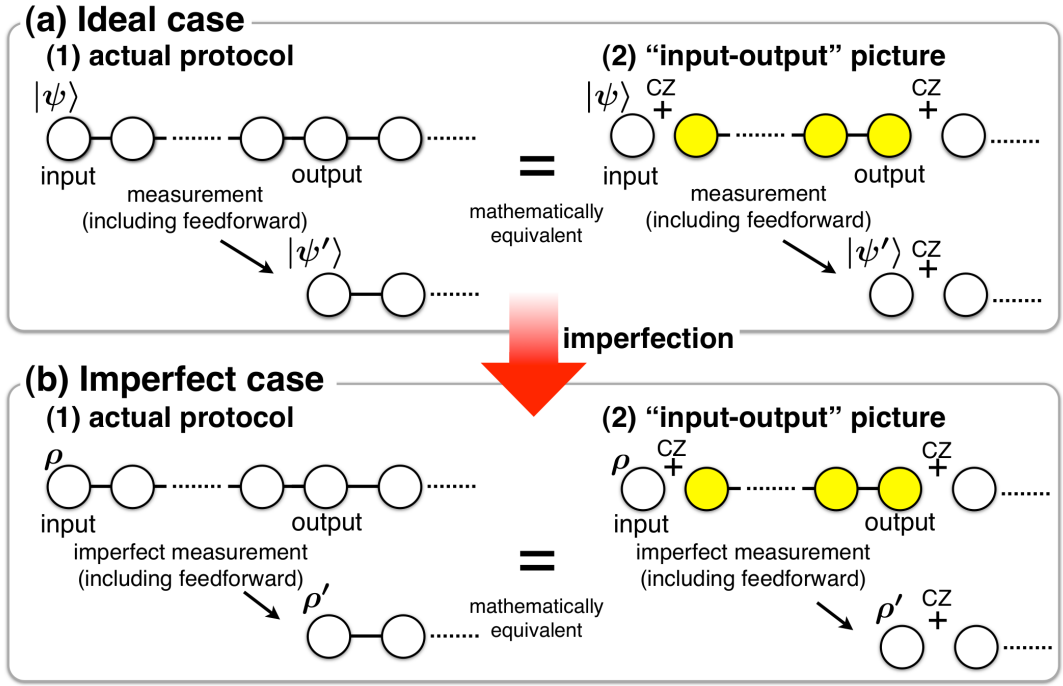}
\end{center}
\caption{
The ``input-output" picture for the one-dimensional cluster state.
} 
\label{hoge1}
\end{figure}

Figure~\ref{hoge1} illustrates
the reason why all physical errors become CPTP maps in the correlation space
of the one-dimensional cluster state.
Let us first consider the ideal case (a) where there is no error.
The actual protocol (a-1) is mathematically equivalent
to the ``input-output" picture (a-2) where the physical input state 
$|\psi\rangle$
is teleported
into the left-edge of the short-length chain (indicated in yellow) and finally 
the physical output state $|\psi'\rangle$ is extracted
from the right-edge of the short-length chain (indicated in yellow).
Since both $|\psi\rangle$ and $|\psi'\rangle$
are physical states,
what is going on 
in the correlation space which maps
the input state $|\psi\rangle$ to the output state $|\psi'\rangle$
can be described by a linear CPTP operation. 
In other words, if we can describe measurement-based
quantum computation with
this ``input-output" picture~\cite{AL06},
the map implemented in the correlation space is guaranteed to be a linear CPTP
operation~\cite{AL06}.

For the cluster state,
this ``input-output" picture also holds even if there is an error~\cite{AL06}:
In the imperfect case, Fig.~\ref{hoge1} (b), let us assume that the input 
state is degraded by an error and becomes a mixed state $\rho$.
However, we can still consider a similar ``input-output" picture (b-2),
which corresponds to the actual protocol (b-1),
and again the physical state $\rho$ is mapped into another physical state
$\rho'$,
which means that what is going on in the correlation space
which maps $\rho$ to $\rho'$ can be
described by a linear CPTP operation.

Note that two special properties of the cluster state
enable such an ``input-output" picture.
First, the one-dimensional 
cluster state can be decomposed into
small pieces of one-dimensional cluster states
by applying nearest-neighbour two-body unitary operations (i.e., C$Z$ gates).
As is shown in Fig.~\ref{hoge1} (a-2) and (b-2),
this property is necessary for
allowing the ``input-output" picture.
Second, the number of qubits that are measured
in order to 
implement a specific gate
does not depend on the measurement results.
In other words, for the cluster state,
a specific gate can be implemented up to Pauli
byproducts at a {\it fixed}
site irrespective of measurement results.
Such a deterministic implementation 
at a fixed site
is necessary for
the deterministic (i.e., trace-preserving)
``output" in the
``input-output" picture,
since 
in the method of Refs.~\cite{AL06},
the ensemble (mixture) of all measurement
results are considered:
If the site where the desired gate operation
is completed depends on the measurement
results,
we cannot ``extract" {\it the same}
output state 
at a {\it fixed} site 
irrespective of measurement results
as is shown in
Fig.~\ref{hoge1} (a-2) and (b-2). 

\begin{figure}[htbp]
\begin{center}
\includegraphics[width=0.5\textwidth]{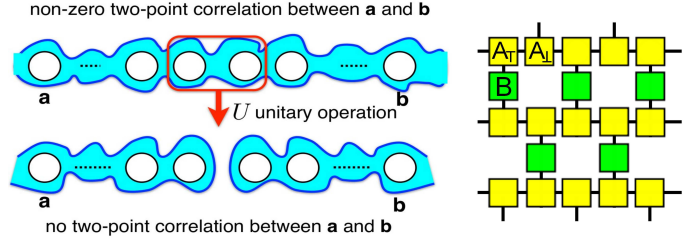}
\end{center}
\caption{
Left:
The one-dimensional AKLT state cannot be decomposed into
short chains.
Right:
The tensor network for the resource state in Ref.~\cite{Caimagnet}.
} 
\label{hoge2}
\end{figure}

On the other hand, 
such an ``input-output"
picture seems to be
impossible for the one-dimensional AKLT state, 
because of the following two reasons:
First, as is shown in Fig.~\ref{hoge2} left, 
no nearest-neighbour two-body 
unitary operation can decompose the one-dimensional AKLT
chain into two chains due to the existence of the non-vanishing two-point
correlation in the AKLT state.
(If the one-dimensional AKLT chain can be decomposed into 
two chains by such a unitary, it contradicts to the well-known
fact that the two-point
correlation is non-vanishing in the AKLT state.)
Second, we cannot deterministically
implement a specific gate at a {\it fixed} site
of the AKLT chain irrespective of measurement
results~\cite{Gross1,Gross2,Brennen}.
In short, the ``input-output"
picture seems to be impossible for the AKLT state.
If we can no longer use the ``input-output" picture,
it is not unreasonable that we have some anomalous
maps in the correlation space 
since the correlation space is not a physical
space but an abstract mathematical space.

\section{Another way of simulating quantum circuits}
\label{sec:pure}

As we have seen in the previous section,
it is not always possible 
for general resource states
to implement a specific gate
at a fixed site of the resource state irrespective of
measurement results.
This fact prohibits a general resource state from
allowing the ``input-output" picture.
One might think that
if we abandon such a deterministic ``output"
at a fixed site
for all measurement results,
and if we just consider a specific history (trajectory)
of measurement results,
we might be able to avoid the emergence of non-CPTP errors.
(Physically, this means that 
we project the system onto a 
pure state at every measurement step.)
If the system is assumed to be error-free,
this ``trajectory method" is another standard way of simulating quantum 
circuits in the correlation space~\cite{Gross1,Gross2,Gross3,Download,Brennen}.
(Note that in this trajectory method, correct
unitary operators can be implemented in the
correlation space if there is no error, although what
we are physically doing are projections, i.e., non-trace-preserving
operations).

However,
we here show that such a natural another way of simulating
quantum circuits does neither work if $d\ge3$.
In other words, we can show the following theorem.

\subsection{Theorem}

{\bf Theorem}:
If $d\ge3$, there exists a single-qudit CPTP error $\mathcal E$ 
which has the following property:
assume that $\mathcal E$ is applied on a single physical qudit of 
$|\Psi(L,R)\rangle_1^N$.
If the measurement ${\mathcal M}_{\theta,\phi}$, Eq.~(\ref{measurement}), 
is performed on 
that affected qudit, 
a non-TP operation is implemented in the correlation space.

{\bf Proof}:
In order to show Theorem, let us assume that  
\begin{eqnarray}
\mbox{There is no such $\mathcal E$}. 
\label{assumption}
\end{eqnarray}
We will see that this assumption leads to
the contradiction that $d\le2$.

First, let us consider 
the state
\begin{eqnarray}
(I^{\otimes N-1}\otimes U_{1\leftrightarrow 2})
|\Psi(L,R)\rangle_1^N,
\label{error1}
\end{eqnarray}
where  
\begin{eqnarray*}
U_{a\leftrightarrow b}\equiv
|a\rangle\langle b|+|b\rangle\langle a|
+I
-|a\rangle\langle a|-|b\rangle\langle b|
\end{eqnarray*}
is the unitary error which exchanges $|a\rangle$ and $|b\rangle$,
and $I$ is the identity operator on a single qudit.
In Eq.~(\ref{error1}), 
the error $U_{1\leftrightarrow2}$ is applied on the first physical
qudit of $|\Psi(L,R)\rangle_1^N$.
If we do the measurement ${\mathcal M}_{\theta,\phi}$ on the first physical
qudit of
Eq.~(\ref{error1}),
and if the measurement result is $|2\rangle$,
Eq.~(\ref{error1})
becomes
\begin{eqnarray}
\frac{1}{\sqrt{f_{N-1}(|L\rangle,A[1]|R\rangle)}}
\|A[1]\|
\sum_{k_2=0}^{d-1}
...
\sum_{k_N=0}^{d-1}
\langle L|A[k_N]...A[k_2]\frac{A[1]}{\|A[1]\|}|R\rangle
|k_N,...,k_2\rangle\otimes|2\rangle.
\label{error1_after}
\end{eqnarray}
In other words, the operation 
\begin{eqnarray*}
|R\rangle\to
\frac{A[1]}{\|A[1]\|}|R\rangle 
\end{eqnarray*}
is implemented in the correlation space.
By the assumption Eq.~(\ref{assumption}),
this operation should work as a TP operation in the correlation space.
Therefore,
\begin{eqnarray}
\frac{A^\dagger[1]}{\|A[1]\|}
\frac{A[1]}{\|A[1]\|}=I.
\label{error1_result}
\end{eqnarray}
By taking $\eta\equiv\|A[1]\|^2$,
\begin{eqnarray}
A^\dagger[1]A[1]=\eta I.
\label{error1_result}
\end{eqnarray}

Second, let us consider the measurement ${\mathcal M}_{\theta,\phi}$ on 
the first physical qudit of
\begin{eqnarray*}
(I^{\otimes N-1}\otimes U_{0\leftrightarrow 2}V^s)
|\Psi(L,R)\rangle_1^N,
\end{eqnarray*}
where  
$s\in\{0,1,...,d-1\}$, 
\begin{eqnarray*}
V\equiv\sum_{p=0}^{d-1}e^{-i\omega p}|p\rangle\langle p|
\end{eqnarray*}
is a unitary phase error,
and
$\omega\equiv 2\pi/d$.
If the measurement result is $|\alpha_{\theta,\phi}\rangle$,
\begin{eqnarray*}
\Big(e^{-2is\omega}\cos\frac{\theta}{2}A[2]
+e^{-i(\phi+s\omega)}\sin\frac{\theta}{2}A[1]
\Big)/\sqrt{\gamma}
\end{eqnarray*}
is implemented in the correlation space,
where
\begin{eqnarray*}
\sqrt{\gamma}\equiv
\Big\|e^{-2is\omega}\cos\frac{\theta}{2}A[2]
+e^{-i(\phi+s\omega)}\sin\frac{\theta}{2}A[1]
\Big\|.
\end{eqnarray*}
By the assumption Eq.~(\ref{assumption}), this should
work as a TP operation in the correlation space.
Therefore,
\begin{eqnarray*}
\gamma I&=&\cos^2\frac{\theta}{2}A^\dagger[2]A[2]+
\sin^2\frac{\theta}{2}A^\dagger[1]A[1]
+\frac{1}{2}
\sin\theta
\Big(e^{-i(\phi-s\omega)}A^\dagger[2]A[1]
+e^{i(\phi-s\omega)}A^\dagger[1]A[2]\Big).
\end{eqnarray*}
By the assumption Eq.~(\ref{assumption0}), 
\begin{eqnarray*}
A^\dagger[2]A[2]=\xi I,
\end{eqnarray*}
where $\xi\equiv\|A[2]\|^2$.
Furthermore, as we have shown, $A^\dagger[1]A[1]=\eta I$ (Eq.~(\ref{error1_result})).
Therefore,
\begin{eqnarray}
\gamma' I=
e^{-i(\phi-s\omega)}A^\dagger[2]A[1]+
e^{i(\phi-s\omega)}A^\dagger[1]A[2],
\label{e3}
\end{eqnarray}
where 
\begin{eqnarray*}
\gamma'\equiv \frac{2}{\sin\theta}\Big(\gamma-\xi\cos^2\frac{\theta}{2}
-\eta\sin^2\frac{\theta}{2}\Big).
\end{eqnarray*}

Finally, let us consider 
the measurement ${\mathcal M}_{\theta,\phi}$ on 
the first physical qudit of
\begin{eqnarray*}
(I^{\otimes N-1}\otimes U_{0\leftrightarrow 1}U_{0\leftrightarrow 2}V^t)
|\Psi(L,R)\rangle_1^N,
\end{eqnarray*}
where $t\in\{0,1,...,d-1\}$.
If the measurement result
is $|\alpha_{\theta,\phi}\rangle$,
\begin{eqnarray*}
\Big(e^{-it\omega}\cos\frac{\theta}{2} A[1]
+e^{-i\phi-2it\omega}\sin\frac{\theta}{2} A[2]\Big)/\sqrt{\delta}
\end{eqnarray*}
is implemented in the correlation space,
where
\begin{eqnarray*}
\sqrt{\delta}\equiv\Big\|e^{-it\omega}\cos\frac{\theta}{2} A[1]
+e^{-i\phi-2it\omega}\sin\frac{\theta}{2} A[2]\Big\|.
\end{eqnarray*}
By the assumption Eq.~(\ref{assumption}), 
this should also work as a TP operation in the correlation
space. Therefore,
\begin{eqnarray}
\delta' I=
e^{i(\phi+t\omega)}A^\dagger[2]A[1]
+e^{-i(\phi+t\omega)}A^\dagger[1]A[2],
\label{e4}
\end{eqnarray}
where 
\begin{eqnarray*}
\delta'\equiv\frac{2}{\sin\theta}\Big(\delta-\xi\sin^2\frac{\theta}{2}
-\eta\cos^2\frac{\theta}{2}\Big).
\end{eqnarray*}

From Eqs.~(\ref{e3}) and (\ref{e4}), 
\begin{eqnarray*}
\epsilon I=
\Big[
e^{-2i(\phi-s\omega)}-e^{2i(\phi+t\omega)}
\Big]A^\dagger[2]A[1],
\end{eqnarray*}
where 
\begin{eqnarray*}
\epsilon\equiv e^{-i(\phi-s\omega)}\gamma'-e^{i(\phi+t\omega)}\delta'.
\end{eqnarray*}

Let us assume that
\begin{eqnarray*}
e^{-2i(\phi-s\omega)}-e^{2i(\phi+t\omega)}\neq0.
\end{eqnarray*}
Then,
\begin{eqnarray*}
\epsilon' I=A^\dagger[2]A[1],
\end{eqnarray*}
where 
\begin{eqnarray*}
\epsilon'\equiv \frac{\epsilon}{
e^{-2i(\phi-s\omega)}-e^{2i(\phi+t\omega)}}.
\end{eqnarray*}
If $\epsilon'=0$,
$A^\dagger[2]A[1]=0$, which means $A[1]=0$ since $A[2]$ is unitary
up to a constant (assumption Eq.~(\ref{assumption0})).
Therefore, $\epsilon'\neq0$.
In this case,
$A[1]=\epsilon'' A[2]$ for certain $\epsilon''\neq0$, 
since $A[2]$ is unitary up to a constant~\cite{exclude}.
Hence 
\begin{eqnarray*}
e^{-2i(\phi-s\omega)}-e^{2i(\phi+t\omega)}=0.
\end{eqnarray*}
This means
\begin{eqnarray}
2\phi+(t-s)\omega=r_{s,t}\pi,
\label{theorem}
\end{eqnarray}
where $r_{s,t}\in\{0,1,2,3,...\}$.
Let us take $t=s=0$.
Then, Eq.~(\ref{theorem}) gives
\begin{eqnarray*}
\phi=r_{0,0}\frac{\pi}{2}~~~(r_{0,0}\in\{0,1,2,...\}).
\end{eqnarray*}
Let us take $s=1$, $t=0$.
Then, Eq.~(\ref{theorem}) gives
\begin{eqnarray*}
\phi=\frac{\pi}{d}+r_{1,0}\frac{\pi}{2}~~~(r_{1,0}\in\{0,1,2,...\}).
\end{eqnarray*}
In order to satisfy these two equations at the same time,
there must exist $r_{0,0}$ and $r_{1,0}$
such that
\begin{eqnarray*}
r_{0,0}\frac{\pi}{2}
=\frac{\pi}{d}+r_{1,0}\frac{\pi}{2}.
\end{eqnarray*}
If $r_{0,0}=r_{1,0}$, then $0=1/d$ which means $d=\infty$.
Therefore
$r_{0,0}\neq r_{1,0}$.
Then we have
\begin{eqnarray*}
d=\frac{2}{r_{0,0}-r_{1,0}}
\le 2,
\end{eqnarray*}
which is the contradiction.~$\blacksquare$

One might think that if 
we rewrite the post-measurement state
Eq.~(\ref{error1_after})
as
\begin{eqnarray*}
\frac{1}{\sqrt{f_{N-1}(|L\rangle,A[1]|R\rangle)}}
\|A[1]|R\rangle\|
\sum_{k_2=0}^{d-1}
...
\sum_{k_N=0}^{d-1}
\langle L|A[k_N]...A[k_2]
\frac{A[1]}
{\|A[1]|R\rangle\|}
|R\rangle
|k_N,...,K_2\rangle\otimes|2\rangle
\end{eqnarray*}
and redefine the operation implemented in the correlation space
as
\begin{eqnarray*}
|R\rangle\to
\frac{A[1]}
{\|A[1]|R\rangle\|}
|R\rangle,
\end{eqnarray*}
the TP-ness is recovered in the correlation space.
However, in this case, the non-lineally appears
unless 
\begin{eqnarray*}
A^\dagger[1]A[1]\propto I,
\end{eqnarray*}
and therefore if we require the linearity in the correlation space,
we obtain the same contradiction.

In short,
if $d\ge3$
not all physical errors on physical qudits
appear as linear CPTP 
errors in the correlation space of pure matrix product 
states~\cite{intuition}.

\subsection{Examples}
Let us consider some
concrete examples which give intuitive understandings
of the above theorem.

In Ref.~\cite{Brennen}, it was shown that 
universal single-qubit unitary rotation
is possible in the correlation space of the one-dimensional 
AKLT chain,
which is a ground state of a gapped
two-body nearest-neighbour
spin-1 Hamiltonian (hence $d=3$).
The matrix product representation of the
one-dimensional AKLT chain is given by 
\begin{eqnarray*}
A[0]&=&X,\\
A[1]&=&XZ,\\
A[2]&=&Z
\end{eqnarray*}
for certain basis $\{|0\rangle,|1\rangle,|2\rangle\}$~\cite{Brennen}.
The measurement 
\begin{eqnarray*}
{\mathcal M}_{\theta,\pi/2}\equiv\Big\{
\cos\frac{\theta}{2}|0\rangle
+i\sin\frac{\theta}{2}|1\rangle,
\sin\frac{\theta}{2}|0\rangle
-i\cos\frac{\theta}{2}|1\rangle,
|2\rangle\Big\}
\end{eqnarray*}
on a single physical qutrit
implements 
$Xe^{-iZ\theta/2}$, 
$XZe^{-iZ\theta/2}$, 
or $Z$,
respectively.
According to Theorem in the previous section, 
not all physical errors can be linear CPTP
maps in the correlation space since $d\ge3$.
In fact, let us consider
the single qutrit unitary error
\begin{eqnarray*}
U\equiv
|2\rangle\frac{\langle0|+\langle1|}{\sqrt{2}}
+\frac{|0\rangle+|1\rangle}{\sqrt{2}}\langle2|
+\frac{|0\rangle-|1\rangle}{\sqrt{2}}\frac{\langle0|-\langle1|}{\sqrt{2}}.
\end{eqnarray*}
If the measurement ${\mathcal M}_{\theta,\pi/2}$ is performed after
the error $U$ and if we obtain the result $|2\rangle$,
the operation $|1\rangle\langle0|$
is implemented in the correlation space. 
Obviously, it is not TP.
The same result is obtained for
the slightly modified version
of the one-dimensional AKLT chain~\cite{Gross1}, 
where
\begin{eqnarray*}
A[0]&=&X,\\
A[1]&=&XZ,\\
A[2]&=&H.
\end{eqnarray*}

In Ref.~\cite{Caimagnet}, it was shown that the unique ground state of a
gapped two-body nearest-neighbour spin-$3/2$ Hamiltonian
with the AKLT and exchange
interactions on the 
two-dimensional octagonal lattice
is a universal resource state 
for the measurement-based quantum computation.
The state is defined by the following tensor network~\cite{Caimagnet}: 
\begin{eqnarray*}
A_{\top}\Big[+\frac{3}{2}\Big]&=&|1\rangle\langle0|\otimes\langle1|,\\
A_{\top}\Big[-\frac{3}{2}\Big]&=&|0\rangle\langle1|\otimes\langle0|,\\
A_{\top}\Big[+\frac{1}{2}\Big]&=&-\frac{1}{\sqrt{3}}\Big(Z\otimes\langle1|
+|1\rangle\langle0|\otimes\langle0|\Big),\\
A_{\top}\Big[-\frac{1}{2}\Big]&=&\frac{1}{\sqrt{3}}\Big(Z\otimes\langle0|
-|0\rangle\langle1|\otimes\langle1|\Big),\\
B\Big[+\frac{3}{2}\Big]&=&|0\rangle\langle1|,\\
B\Big[-\frac{3}{2}\Big]&=&-|1\rangle\langle0|,\\
B\Big[+\frac{1}{2}\Big]&=&|1\rangle\langle1|,\\
B\Big[-\frac{1}{2}\Big]&=&-|0\rangle\langle0|,
\end{eqnarray*}
(see Fig.~\ref{hoge2} right).
$A_\perp$'s are defined in the same way.
Each horizontal line works as a single-qubit wire.
Two nearest-neighbour horizontal chains are decoupled by measuring 
sites B in the $z$-basis.
Before starting the computation,
the filtering operation 
$\{F,F'\}$,
where
\begin{eqnarray*}
F&\equiv&
\frac{1}{\sqrt{3}}|3/2\rangle\langle3/2|
+\frac{1}{\sqrt{3}}|-3/2\rangle\langle-3/2|
+|1/2\rangle\langle1/2|
+|-1/2\rangle\langle-1/2|\\
F'&\equiv&
\sqrt{\frac{2}{3}}|3/2\rangle\langle3/2|
+\sqrt{\frac{2}{3}}|-3/2\rangle\langle-3/2|,
\end{eqnarray*}
is applied on each site A.
Let us assume that the filtering is succeeded (i.e.,
$F$ is realized)
and a site B is projected onto $|3/2\rangle$.
Then,
the measurement
\begin{eqnarray*}
{\mathcal M}_{\pi/2,\phi}\equiv\Big\{
\frac{1}{\sqrt{2}}
\Big(|1/2\rangle\pm e^{i\phi}|-3/2\rangle\Big),
|-1/2\rangle,
|+3/2\rangle
\Big\}
\end{eqnarray*}
implements 
$ZXe^{iZ\phi/2}$,
$Xe^{iZ\phi/2}$,
or
$Z$,
respectively.
(The result $|3/2\rangle$ does not occur.)
If the error which exchanges
$|1/2\rangle$ and $|-1/2\rangle$
occurs and if the result of the measurement $\mathcal M_{\pi/2,\phi}$
is $|-1/2\rangle$, the operation 
$|1\rangle\langle0|$, which is not TP, is implemented 
in the correlation space.


\section{Conclusion}
\label{sec:conclusion}

In this paper, we have studied how physical errors on a physical qudit
appear in the correlation space of general resource states.
We have shown that 
the results~\cite{ND05,AL06} for the cluster state cannot be
directly applied to
general resource states, such the AKLT state.
We have also shown that if $d\ge3$ not all physical errors
can be linear CPTP errors in the correlation space of 
pure matrix product states.
These results suggest that the application of
the theories of fault-tolerant quantum circuits
to the correlation space of general resource states
is not so straightforward.

\acknowledgements
TM and KF acknowledge supports by ANR (StatQuant, JC07 07205763)
and MEXT
(Grant-in-Aid for Scientific Research
on Innovative Areas 20104003), respectively.

\appendix

\section{Tricluster state without error}
\label{App:tri1}
Let us consider the tricluster state~\cite{tricluster},
where $d=6$ and
\begin{eqnarray*}
A[0]&=&|+\rangle\langle0|,\\
A[1]&=&|-\rangle\langle1|,\\
A[2]&=&|-\rangle\langle0|,\\
A[3]&=&|+\rangle\langle1|,\\
A[4]&=&|+\rangle\langle1|,\\
A[5]&=&|-\rangle\langle0|.
\end{eqnarray*}
The measurement
\begin{eqnarray*}
|\theta_0\rangle&\equiv&\frac{1}{\sqrt{2}}\Big(|0\rangle+e^{i\theta}|1\rangle\Big)\\
|\theta_1\rangle&\equiv&\frac{1}{\sqrt{2}}\Big(|0\rangle-e^{i\theta}|1\rangle\Big)\\
|\theta_2\rangle&\equiv&\frac{1}{\sqrt{2}}\Big(|2\rangle+e^{i\theta}|3\rangle\Big)\\
|\theta_3\rangle&\equiv&\frac{1}{\sqrt{2}}\Big(|2\rangle-e^{i\theta}|3\rangle\Big)\\
|\theta_4\rangle&\equiv&\frac{1}{\sqrt{2}}\Big(|4\rangle+e^{-i\theta}|5\rangle\Big)\\
|\theta_5\rangle&\equiv&\frac{1}{\sqrt{2}}\Big(|4\rangle-e^{-i\theta}|5\rangle\Big)
\end{eqnarray*}
is performed in order to implement
\begin{eqnarray*}
|\theta_0\rangle&:&J(\theta)\\
|\theta_1\rangle&:&XJ(\theta)\\
|\theta_2\rangle&:&ZJ(\theta)\\
|\theta_3\rangle&:&ZXJ(\theta)\\
|\theta_4\rangle&:&ZJ(\theta)\\
|\theta_5\rangle&:&ZXJ(\theta).
\end{eqnarray*}

First, we measure the first physical qudit in the 
$\{|\theta_{s_1}\rangle\}$
basis.
Then we obtain
\begin{eqnarray*}
\frac{1}{2f_N(|L\rangle,|R\rangle)}
\sum_{s_1=0}^5W(
X^{p(s_1)}
Z^{q(s_1)}
J(\theta)
|R\rangle)_2\otimes
|\theta_{s_1}\rangle\langle\theta_{s_1}|\otimes m(s_1),
\end{eqnarray*}
where 
\begin{eqnarray*}
p(s)&\equiv& \delta_{s,1}+\delta_{s,3}+\delta_{s,5}\\
q(s)&\equiv& \delta_{s,2}+\delta_{s,3}+\delta_{s,4}+\delta_{s,5}.
\end{eqnarray*}
If we trace out the first physical qudit, we obtain
\begin{eqnarray*}
\frac{1}{2f_N(|L\rangle,|R\rangle)}
\sum_{s_1=0}^5W(
X^{p(s_1)}
Z^{q(s_1)}
J(\theta)|R\rangle)_2\otimes
m(s_1).
\end{eqnarray*}

Second, if we measure the second physical qudit in
the basis
\begin{eqnarray}
|s_1,\phi_0\rangle&=&\frac{1}{\sqrt{2}}\Big(|0\rangle+(-1)^{q(s_1)}e^{i\phi(-1)^{p(s_1)}}|1\rangle\Big)\nonumber\\
|s_1,\phi_1\rangle&=&\frac{1}{\sqrt{2}}\Big(|0\rangle-(-1)^{q(s_1)}e^{i\phi(-1)^{p(s_1)}}|1\rangle\Big)\nonumber\\
|s_1,\phi_2\rangle&=&\frac{1}{\sqrt{2}}\Big(|2\rangle+(-1)^{q(s_1)}e^{i\phi(-1)^{p(s_1)}}|3\rangle\Big)\nonumber\\
|s_1,\phi_3\rangle&=&\frac{1}{\sqrt{2}}\Big(|2\rangle-(-1)^{q(s_1)}e^{i\phi(-1)^{p(s_1)}}|3\rangle\Big)\nonumber\\
|s_1,\phi_4\rangle&=&\frac{1}{\sqrt{2}}\Big(|4\rangle+(-1)^{q(s_1)}e^{i\phi(-1)^{p(s_1)}}|5\rangle\Big)\nonumber\\
|s_1,\phi_5\rangle&=&\frac{1}{\sqrt{2}}\Big(|4\rangle-(-1)^{q(s_1)}e^{i\phi(-1)^{p(s_1)}}|5\rangle\Big)
\label{tricluster_basis}
\end{eqnarray}
and trace out the measured qudit,
we obtain
\begin{eqnarray*}
&&
\frac{1}{2^2f_N(|L\rangle,|R\rangle)}
\sum_{s_2=0}^5
\sum_{s_1=0}^5
W(
X^{p(s_2)}
Z^{q(s_2)}
X^{q(s_1)}
J((-1)^{p(s_1)}\phi)
X^{p(s_1)}
Z^{q(s_1)}
J(\theta)|R\rangle)_3
\otimes
m(s_1)
\otimes
m(s_2)\\
&=&
\frac{1}{2^2f_N(|L\rangle,|R\rangle)}
\sum_{s_2=0}^5
\sum_{s_1=0}^5
W(
X^{p(s_2)}
Z^{q(s_2)}
Z^{p(s_1)}
J(\phi)
J(\theta)|R\rangle)_3
\otimes
m(s_1)
\otimes
m(s_2).
\end{eqnarray*}

Third, if we measure the third physical qudit
in the basis 
\begin{eqnarray}
|s_1,s_2,\eta_0\rangle&=&\frac{1}{\sqrt{2}}\Big(|0\rangle+(-1)^{p(s_1)+q(s_2)}
e^{i\eta(-1)^{p(s_2)}}|1\rangle\Big)\nonumber\\
|s_1,s_2,\eta_1\rangle&=&\frac{1}{\sqrt{2}}\Big(|0\rangle-(-1)^{p(s_1)+q(s_2)}
e^{i\eta(-1)^{p(s_2)}}|1\rangle\Big)\nonumber\\
|s_1,s_2,\eta_3\rangle&=&\frac{1}{\sqrt{2}}\Big(|2\rangle+(-1)^{p(s_1)+q(s_2)}
e^{i\eta(-1)^{p(s_2)}}|3\rangle\Big)\nonumber\\
|s_1,s_2,\eta_4\rangle&=&\frac{1}{\sqrt{2}}\Big(|2\rangle-(-1)^{p(s_1)+q(s_2)}
e^{i\eta(-1)^{p(s_2)}}|3\rangle\Big)\nonumber\\
|s_1,s_2,\eta_5\rangle&=&\frac{1}{\sqrt{2}}\Big(|4\rangle+(-1)^{p(s_1)+q(s_2)}
e^{i\eta(-1)^{p(s_2)}}|5\rangle\Big)\nonumber\\
|s_1,s_2,\eta_6\rangle&=&\frac{1}{\sqrt{2}}\Big(|4\rangle-(-1)^{p(s_1)+q(s_2)}
e^{i\eta(-1)^{p(s_2)}}|5\rangle\Big)
\label{tricluster_basis2}
\end{eqnarray}
and trace out the measured third qudit,
we obtain
\begin{eqnarray*}
&&
\frac{1}{2^3f_N(|L\rangle,|R\rangle)}
\sum_{s_3=0}^5
\sum_{s_2=0}^5
\sum_{s_1=0}^5\\
&&W(
X^{p(s_3)}
Z^{q(s_3)}
X^{p(s_1)+q(s_2)}
J((-1)^{p(s_2)}\eta)
X^{p(s_2)}
Z^{q(s_2)}
Z^{p(s_1)}
J(\phi)
J(\theta)|R\rangle)_4
\otimes
m(s_1)
\otimes
m(s_2)
\otimes
m(s_3)\\
&=&
\frac{1}{2^3f_N(|L\rangle,|R\rangle)}
\sum_{s_3=0}^5
\sum_{s_2=0}^5
\sum_{s_1=0}^5
W(
X^{p(s_3)}
Z^{q(s_3)}
Z^{p(s_2)}
J(\eta)
J(\phi)
J(\theta)|R\rangle)_4
\otimes
m(s_1)
\otimes
m(s_2)
\otimes
m(s_3)\\
&=&
\frac{1}{2^3f_N(|L\rangle,|R\rangle)}
\sum_{s_3=0}^5
\sum_{s_2=0}^5
W(
X^{p(s_3)}
Z^{q(s_3)}
Z^{p(s_2)}
J(\eta)
J(\phi)
J(\theta)|R\rangle)_4
\otimes
m(s_2)
\otimes
m(s_3).
\end{eqnarray*}

\section{Tricluster state with error}
\label{App:tri2}
Let us assume that a CPTP error occurs on the first physical qudit.
If we measure the first physical qudit in the 
$\{|\theta_{s_1}\rangle\}$
basis, we obtain
\begin{eqnarray*}
\frac{1}{f_N(|L\rangle,|R\rangle)}
\sum_{s_1=0}^5\sum_jW(E_{j,s_1}|R\rangle)_2
\otimes
|\theta_{s_1}\rangle\langle\theta_{s_1}|\otimes m(s_1).
\end{eqnarray*}
By tracing out the first physical qudit, we obtain
\begin{eqnarray*}
\frac{1}{f_N(|L\rangle,|R\rangle)}
\sum_{s_1=0}^5\sum_jW(E_{j,s_1}|R\rangle)_2
\otimes
m(s_1).
\end{eqnarray*}

Second, if we measure the second physical qudit in
the basis Eq.~(\ref{tricluster_basis})
and trace out the measured qudit,
we obtain
\begin{eqnarray*}
&&
\frac{1}{2f_N(|L\rangle,|R\rangle)}
\sum_{s_2=0}^5
\sum_{s_1=0}^5
\sum_j
W(
X^{p(s_2)}
Z^{q(s_2)}
X^{q(s_1)}
J((-1)^{p(s_1)}\phi)
E_{j,s_1}|R\rangle)_3
\otimes
m(s_1)
\otimes
m(s_2).
\end{eqnarray*}

Third, if we measure the third physical qudit
in the basis Eq.~(\ref{tricluster_basis2})
and trace out the measured third qudit,
we obtain
\begin{eqnarray*}
&&
\frac{1}{2^2f_N(|L\rangle,|R\rangle)}
\sum_{s_3=0}^5
\sum_{s_2=0}^5
\sum_{s_1=0}^5
\sum_j\\
&&W(
X^{p(s_3)}
Z^{q(s_3)}
X^{p(s_1)+q(s_2)}
J((-1)^{p(s_2)}\eta)
X^{p(s_2)}
Z^{q(s_2)}
X^{q(s_1)}
J((-1)^{p(s_1)}\phi)
E_{j,s_1}|R\rangle)_4
\otimes
m(s_1)
\otimes
m(s_2)
\otimes
m(s_3)\\
&=&
\frac{1}{2^2f_N(|L\rangle,|R\rangle)}
\sum_{s_3=0}^5
\sum_{s_2=0}^5
\sum_{s_1=0}^5
\sum_j\\
&&W(
X^{p(s_3)}
Z^{q(s_3)}
Z^{p(s_2)}
X^{p(s_1)}
J(\eta)
X^{q(s_1)}
J((-1)^{p(s_1)}\phi)
E_{j,s_1}|R\rangle)_4
\otimes
m(s_1)
\otimes
m(s_2)
\otimes
m(s_3).
\end{eqnarray*}
If we trace out the first record,
\begin{eqnarray*}
\frac{1}{2^2f_N(|L\rangle,|R\rangle)}
\sum_{s_3=0}^5
\sum_{s_2=0}^5
\sum_{s_1=0}^5
\sum_j
W(
X^{p(s_3)}
Z^{q(s_3)}
Z^{p(s_2)}
X^{p(s_1)}
J(\eta)
X^{q(s_1)}
J((-1)^{p(s_1)}\phi)
E_{j,s_1}|R\rangle)_4
\otimes
m(s_2)
\otimes
m(s_3).
\end{eqnarray*}

Thus the map
\begin{eqnarray*}
|R\rangle\langle R|\to
\sum_{s_1=0}^{5}
\sum_j
X^{p(s_1)}
J(\eta)X^{q(s_1)}
J((-1)^{p(s_1)}\phi)
E_{j,s_1}|R\rangle
\langle R|E_{j,s_1}^\dagger
J^\dagger((-1)^{p(s_1)}\phi)
X^{q(s_1)}
J^\dagger(\eta)
X^{p(s_1)},
\end{eqnarray*}
which is obviously TP, is implemented.

\section{Calculation of $|U_{p,q}^r|$}
Let us define
\begin{eqnarray*}
U_{p,q}^r\equiv
\Big\{
(s_1,...,s_r)\in\{0,1,2\}^{\times r}~\Big|~
f(s_1,...,s_r)=p~\mbox{and}~g(s_1,...,s_r)=q
\Big\}.
\end{eqnarray*}
First,
\begin{eqnarray*}
|U_{0,0}^2|&=&3\\
|U_{0,1}^2|&=&2\\
|U_{1,0}^2|&=&2\\
|U_{1,1}^2|&=&2.
\end{eqnarray*}
Second,
\begin{eqnarray*}
|U_{p,q}^r|=
|U_{p\oplus1,q}^{r-1}|
+|U_{p,q\oplus1}^{r-1}|
+|U_{p\oplus1,q\oplus1}^{r-1}|
\end{eqnarray*}
for all $r$.
Therefore,
\begin{eqnarray*}
|U_{0,1}^r|-|U_{1,0}^r|&=&
|U_{1,1}^{r-1}|
+|U_{0,0}^{r-1}|
+|U_{1,0}^{r-1}|
-|U_{0,0}^{r-1}|
-|U_{1,1}^{r-1}|
-|U_{0,1}^{r-1}|\\
&=&
|U_{1,0}^{r-1}|
-|U_{0,1}^{r-1}|\\
&=&
-(
|U_{0,1}^{r-1}|
-|U_{1,0}^{r-1}|
)\\
&=&
(-1)^{r-2}(
|U_{0,1}^2|
-|U_{1,0}^2|
)\\
&=&0
\end{eqnarray*}
and
\begin{eqnarray*}
|U_{0,1}^r|-|U_{1,1}^r|&=&
|U_{1,1}^{r-1}|
+|U_{0,0}^{r-1}|
+|U_{1,0}^{r-1}|
-|U_{0,1}^{r-1}|
-|U_{1,0}^{r-1}|
-|U_{0,0}^{r-1}|\\
&=&
|U_{1,1}^{r-1}|
-|U_{0,1}^{r-1}|\\
&=&
-(
|U_{0,1}^{r-1}|
-|U_{1,1}^{r-1}|
)\\
&=&
(-1)^{r-2}(
|U_{0,1}^2|
-|U_{1,1}^2|
)\\
&=&0,
\end{eqnarray*}
which mean
\begin{eqnarray*}
|U_{1,1}^r|=|U_{0,1}^r|=|U_{1,0}^r|
\end{eqnarray*}
for all $r$.

Note that
\begin{eqnarray*}
|U_{0,0}^r|-|U_{0,1}^r|
&=&
3|U_{0,1}^{r-1}|-(2|U_{0,1}^{r-1}|+|U_{0,0}^{r-1}|)\\
&=&
|U_{0,1}^{r-1}|-|U_{0,0}^{r-1}|\\
&=&
-(|U_{0,0}^{r-1}|-|U_{0,1}^{r-1}|)\\
&=&
(-1)^{r-2}(|U_{0,0}^{2}|-|U_{0,1}^{2}|)\\
&=&
(-1)^{r-2}.
\end{eqnarray*}
and
\begin{eqnarray*}
|U_{0,0}^r|+3|U_{0,1}^r|
&=&
3|U_{0,1}^{r-1}|+3(2|U_{0,1}^{r-1}|+|U_{0,0}^{r-1}|)\\
&=&
3|U_{0,0}^{r-1}|+9|U_{0,1}^{r-1}|\\
&=&
3(|U_{0,0}^{r-1}|+3|U_{0,1}^{r-1}|)\\
&=&
3^{r-2}(|U_{0,0}^{2}|+3|U_{0,1}^{2}|)\\
&=&
3^r.
\end{eqnarray*}
Therefore,
\begin{eqnarray*}
|U_{0,1}^r|&=&\frac{1}{4}(3^r-(-1)^r)\\
|U_{0,0}^r|&=&\frac{1}{4}(3^r+3(-1)^r).
\end{eqnarray*}

\section{Calculation of $|S_{p,q}^r|$}

Let us define
\begin{eqnarray*}
S_{p,q}^r\equiv
\Big\{
(s_1,...,s_r)\in\{0,1,2\}^{\times r}\setminus(2,...,2)~\Big|~
f(s_1,...,s_r)=p~\mbox{and}~g(s_1,...,s_r)=q
\Big\}.
\end{eqnarray*}
First,
\begin{eqnarray*}
|S_{0,0}^2|&=&2\\
|S_{0,1}^2|&=&2\\
|S_{1,0}^2|&=&2\\
|S_{1,1}^2|&=&2.
\end{eqnarray*}
Second,
\begin{eqnarray*}
|S_{p,q}^r|=
|U_{p\oplus1,q}^{r-1}|
+|U_{p\oplus1,q\oplus1}^{r-1}|
+|S_{p,q\oplus1}^{r-1}|.
\end{eqnarray*}
Therefore,
\begin{eqnarray*}
|S_{0,0}^r|-|S_{0,1}^r|
&=&
|U_{1,0}^{r-1}|
+|U_{1,1}^{r-1}|
+|S_{0,1}^{r-1}|
-|U_{1,1}^{r-1}|
-|U_{1,0}^{r-1}|
-|S_{0,0}^{r-1}|
\\
&=&
-(|S_{0,0}^{r-1}|-|S_{0,1}^{r-1}|)\\
&=&
(-1)^{r-2}(|S_{0,0}^2|-|S_{0,1}^2|)\\
&=&0.
\end{eqnarray*}
and
\begin{eqnarray*}
|S_{1,0}^r|-|S_{1,1}^r|
&=&
|U_{0,0}^{r-1}|
+|U_{0,1}^{r-1}|
+|S_{1,1}^{r-1}|
-|U_{0,1}^{r-1}|
-|U_{0,0}^{r-1}|
-|S_{1,0}^{r-1}|\\
&=&
-(|S_{1,0}^{r-1}|-|S_{1,1}^{r-1}|)\\
&=&
(-1)^{r-2}(|S_{1,0}^2|-|S_{1,1}^2|)\\
&=&0.
\end{eqnarray*}
Hence
\begin{eqnarray*}
|S_{0,0}^r|
&=&
\frac{3^{r-1}-(-1)^{r-1}}{2}+|S_{0,0}^{r-1}|\\
&=&
\frac{3^{r-1}-(-1)^{r-1}}{2}+...+\frac{3^2-(-1)^2}{2}+|S_{0,0}^2|\\
&=&
\frac{3^{r-1}+...+3^2}{2}-\frac{
(-1)^{r-1}+...+(-1)^2}{2}+2\\
&=&
\frac{3^r-3^2}{4}
+\frac{
(-1)^r-(-1)^2}{4}+2.
\end{eqnarray*}
and
\begin{eqnarray*}
|S_{1,0}^r|
&=&
\frac{3^{r-1}+(-1)^{r-1}}{2}+|S_{1,0}^{r-1}|\\
&=&
\frac{3^{r-1}+(-1)^{r-1}}{2}+...+\frac{3^2+(-1)^2}{2}+|S_{1,0}^2|\\
&=&
\frac{3^{r-1}+...+3^2}{2}+\frac{
(-1)^{r-1}+...+(-1)^2}{2}+2\\
&=&
\frac{3^r-3^2}{4}
-\frac{
(-1)^r-(-1)^2}{4}+2.
\end{eqnarray*}

\section{Calculation of $|T_{p,q}^{r,i}|$}
\label{app:T}
It is easy to see that
\begin{eqnarray*}
|T_{p,q}^{r,0}|&=&|U_{p\oplus1,q}^{r-1}|\\
|T_{p,q}^{r,1}|&=&|U_{p\oplus1,q\oplus1}^{r-1}|\\
|T_{p,q}^{r,2}|&=&|S_{p,q\oplus1}^{r-1}|,
\end{eqnarray*}
where
\begin{eqnarray*}
U_{p,q}^r\equiv
\Big\{(s_1,...,s_r)\in\{0,1,2\}^{\times r}~\Big|~
f(s_1,...,s_r)=p~\mbox{and}~g(s_1,...,s_r)=q\Big\}.
\end{eqnarray*}

Since
\begin{eqnarray*}
|T_{0,q}^{r,0}|&=&|U_{1,q}^{r-1}|
=\frac{1}{4}(3^{r-1}-(-1)^{r-1})\\
|T_{0,q}^{r,1}|&=&|U_{1,q\oplus1}^{r-1}|
=\frac{1}{4}(3^{r-1}-(-1)^{r-1})\\
|T_{0,q}^{r,2}|&=&|S_{0,q\oplus1}^{r-1}|
=\frac{1}{4}(3^{r-1}-3^2+(-1)^{r-1}-(-1)^2)+2,
\end{eqnarray*}
we obtain
$|T_{0,q}^{r,0}|=|T_{0,q}^{r,1}|$
and
\begin{eqnarray*}
|T_{0,q}^{r,0}|-|T_{0,q}^{r,2}|&=&
-\frac{(-1)^{r-1}}{2}+\frac{3^2}{4}+\frac{1}{4}-2\\
&=&
\frac{1-(-1)^{r-1}}{2}\\
&=&
\Big\{
\begin{array}{ll}
0&(r=\mbox{odd})\\
1&(r=\mbox{even}).
\end{array}
\end{eqnarray*}

Since
\begin{eqnarray*}
|T_{1,0}^{r,0}|&=&|U_{0,0}^{r-1}|=\frac{1}{4}(3^{r-1}+3(-1)^{r-1})\\
|T_{1,0}^{r,1}|&=&|U_{0,1}^{r-1}|=\frac{1}{4}(3^{r-1}-(-1)^{r-1})\\
|T_{1,0}^{r,2}|&=&|S_{1,1}^{r-1}|=\frac{3^{r-1}-3^2}{4}
-\frac{(-1)^{r-1}-(-1)^2}{4}+2,
\end{eqnarray*}
we obtain
\begin{eqnarray*}
|T_{1,0}^{r,0}|-|T_{1,0}^{r,1}|=(-1)^{r-1},
\end{eqnarray*}
\begin{eqnarray*}
|T_{1,0}^{r,0}|-|T_{1,0}^{r,2}|&=&
\frac{3^2}{4}+(-1)^{r-1}-\frac{1}{4}-2\\
&=&(-1)^{r-1},
\end{eqnarray*}
and
\begin{eqnarray*}
|T_{1,0}^{r,1}|-|T_{1,0}^{r,2}|&=&
\frac{3^2}{4}-\frac{1}{4}-2\\
&=&0.
\end{eqnarray*}

Since
\begin{eqnarray*}
|T_{1,1}^{r,0}|&=&|U_{0,1}^{r-1}|=\frac{1}{4}(3^{r-1}-(-1)^{r-1})\\
|T_{1,1}^{r,1}|&=&|U_{0,0}^{r-1}|=\frac{1}{4}(3^{r-1}+3(-1)^{r-1})\\
|T_{1,1}^{r,2}|&=&|S_{1,0}^{r-1}|=\frac{3^{r-1}-3^2}{4}
-\frac{(-1)^{r-1}-(-1)^2}{4}+2,
\end{eqnarray*}
we obtain 
\begin{eqnarray*}
|T_{1,1}^{r,0}|-|T_{1,1}^{r,1}|=-(-1)^{r-1},
\end{eqnarray*}
\begin{eqnarray*}
|T_{1,1}^{r,0}|-|T_{1,1}^{r,2}|&=&
\frac{3^2}{4}-\frac{1}{4}-2\\
&=&0,
\end{eqnarray*}
and
\begin{eqnarray*}
|T_{1,1}^{r,1}|-|T_{1,1}^{r,2}|&=&
\frac{3^2}{4}+(-1)^{r-1}-\frac{1}{4}-2\\
&=&(-1)^{r-1}.
\end{eqnarray*}


\begin{thebibliography}{00}
\bibitem{recently}
Recently, it was shown that topologically-protected
measurement-based quantum computation~\cite{topological1,topological2}
is possible on the thermal equilibrium state
of a nearest-neighbour two-body Hamiltonian
with spin-3/2 and spin-2 particles~\cite{Li}
or solely with spin-3/2 particles~\cite{FMtopo}.
\bibitem{exclude}
We do not consider the case where
$A[1]\propto A[2]$,
since in this case we can reduce the dimension $d$ to $d-1$
by redefining the basis of the two-dimensional subspace
spanned by $|1\rangle$
and $|2\rangle$.
\bibitem{intuition}
This result is reasonable
since we physically do
a non-linear (or non-TP) operation.
If we consider this fact, it is surprising
that 
linear CPTP operations are implemented by 
doing non-linear (or non-TP) physical operations
on several resource states such as the cluster state
when physical operations are perfect!
\end{thebibliography}
\end{document}